\documentclass[showpacs,preprintnumbers,amsmath,amssymb,prl,epsf]{revtex4}
\usepackage{graphicx}
\newcommand{\be}{\begin{equation}}
\newcommand{\ee}{\end{equation}}
\newcommand{\bea}{\vspace{0.25cm}\begin{eqnarray}}
\newcommand{\eea}{\end{eqnarray}}

\begin{document}


\title{
Improved implementation of nonclassicality test for a single
particle }
\author{Giorgio Brida$^{1}$, Ivo Pietro Degiovanni$^{1*}$,  Marco
Genovese$^{1}$,\\
Fabrizio Piacentini$^{1}$, Valentina Schettini$^{1}$, Nicolas
Gisin$^{2},$
\\ Sergey V. Polyakov$^{3}$, and Alan Migdall$^{3}$}

\address{$^{1}$Istituto Nazionale di Ricerca Metrologica, Strada
delle Cacce 91, 10135 Torino, Italy \\
$^{2}$Group of Applied Physics, University of Geneva, CH-1211 Geneva 4, Switzerland\\
$^{3}$Optical Technology Division, National Institute of Standards
and Technology, 100 Bureau Drive, Gaithersburg, MD 20899-8441 and
Joint Quantum Institute, Univ. of Maryland, College Park, MD 20742\\
$^{*}$Corresponding author: i.degiovanni@inrim.it}



\begin{abstract}
Recently  a test of nonclassicality for a single qubit was proposed
[R. Alicki and N. Van Ryn, J. Phys. A: Math. Theor. \textbf{41},
062001 (2008)]. We present an optimized experimental realization of
this test leading to a 46 standard deviation violation of
classicality. This factor of 5 improvement over our previous result
was achieved by moving from the infrared to the visible where we can
take advantage of higher efficiency and lower noise photon
detectors.
\end{abstract}

 \maketitle

\section{Introduction}

A simple test of nonclassicality at the single qubit level was
proposed \cite{al, al2} to show that some quantum states in a two
dimensional Hilbert space cannot be classical. This test looks very
appealing for its simplicity compared to other tests of quantumness
(see \cite{pr} and references therein) and could represent a useful
tool for various applications in the fields of quantum information,
fundamental quantum optics, foundations of quantum mechanics, etc.

As this is a test of single-particle states, it is not relevant to
the question of locality, rather it is a more fundamental test of
nonclassicality with respect to the possibility of an underlying
hidden variable theory (HVT). Furthermore, we understand that it
does not apply to every conceivable HVT like Bell's inequalities,
but only to a restricted class of HVTs \cite{oeske}. We also note
that the precise identification of this class and whether and if it
maps to any physical system at all remains to be determined. In
addition, another question has been raised about this test by
Zukowski \cite{zuk08}, who suggests that the Alicki's classicality
criterion is equivalent to the von Neumann theorem.

The criterion for classicality in Alicki's  model is summarized by
the following statement: for any pair of observables $\widehat{A}$
and $\widehat{B}$ that satisfy the condition
\begin{equation}
\langle \widehat{B}\rangle
> \langle \widehat{A} \rangle > 0
\end{equation}
for all states of the system, it must be always true that
\begin{equation}\label{classCond}
\langle \widehat{B}^{2}\rangle
> \langle \widehat{A}^{2}\rangle.
\end{equation}

For quantum systems, one can find pairs of observables
$\widehat{A},\widehat{B}$ such that the minimum eigenvalue of
$\widehat{B} - \widehat{A}$ (minimized over all possible states) is
greater than zero, while for certain quantum states
\begin{equation}
\langle \widehat{B}^2 \rangle < \langle \widehat{A}^2 \rangle .
\label{qubitbellViol}
\end{equation}
This sharp difference between classicality and nonclassicality in
Alicki's  model can be tested experimentally at the single-qubit
level \cite{oeske}.

One possible pair of operators $\widehat{A}$ and $\widehat{B}$ are
of the form \cite{oeske}
\begin{eqnarray}
\widehat{A}&=& a \frac{\mathbf{1}+\widehat{Z}}{2} \\
\widehat{B}&=& b \frac{\mathbf{1}+r \cos \beta \widehat{Z}+ r \sin
\beta \widehat{X}}{2},
\end{eqnarray}
where $\widehat{Z}$, $\widehat{X}$ are Pauli matrices. To ensure the
positivity of $\widehat{A}$ and $\widehat{B}$, we assume $a > 0$, $b
> 0$, and $0 \leq r \leq 1$ for $a$, $b$, and $r$ real.

For it to be true that given $\langle \widehat{B}\rangle
> \langle \widehat{A}\rangle$, at least one state can be found such that $\langle \widehat{B}^{2}\rangle
< \langle \widehat{A}^{2} \rangle$ (i.e., the condition for Alicki's
nonclassicality), the minimum of the eigenvalue of $ \widehat{B} -
\widehat{A}$ should be positive while $\widehat{B}^{2} -
\widehat{A}^{2}$ should be negative for at least one eigenvalue. It
can easily be shown that these conditions correspond to
\begin{equation}\label{condvio}
\frac{1-r^{2}}{2\sqrt{1+r^{2}-2 r \cos\beta }}
<\frac{a}{b}<\frac{1-r^{2}}{2(1- r \cos\beta)}.
\end{equation}

In this work we present a second experimental realization of
Alicki's test on a single qubit after the one of Ref. \cite{oeske}.
We exploit a simplified and more efficient scheme that achieves a
larger violation of the ãclassicalä inequality in Eq.
(\ref{classCond}) and does so with better uncertainty (a 5x
improvement over the previous result).

\begin{figure}
\begin{center}
\includegraphics[height=.3\textheight]{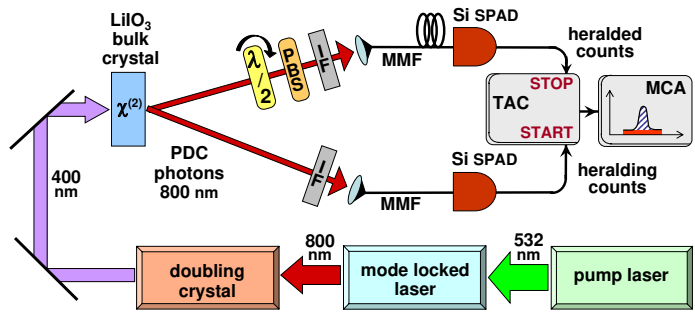}
\end{center} \caption{Experimental setup. A PDC heralded
single-photon source generates pairs of photons at 800 nm in a
LiIO$_{3}$ bulk crystal pumped by a 400 nm mode locked laser. The
counted photon on the heralding arm announces the presence of its
twin on the heralded arm. The coincidence apparatus based on a Time
to Amplitude Converter (TAC) and a Multi Channel Analyzer (MCA) is
used to discriminate coincidences due to PDC twin-photon from the
background. The heralded photons are sent to the measurement
apparatus designed to evaluate the observables
$\langle\widehat{A}\rangle$, $\langle\widehat{A}^{2}\rangle$,
$\langle\widehat{B}\rangle$, and $\langle\widehat{B}^{2}\rangle$.}
\label{setup2}
\end{figure}

\section{The  Experiment}

The quantum objects we use to implement this test are linearly
polarized single-photons ($|\Psi \rangle= \cos\psi |H
\rangle+\sin\psi |V \rangle$) produced by a heralded single-photon
source based on parametric down conversion (PDC). The main
difference of this experiment with respect to the previous one is
that in this case both the heralded and the heralding photons are in
the visible, while in the previous case the heralded photon was at a
telecom wavelength \cite{oeske}. This allows us to use more
efficient and lower noise detection systems that significantly
reduce the experimental uncertainty. We also note that in this
experiment measuring the two operators requires manually changing
the waveplate angle rather than switching between the two
measurements in an automated fashion, an inconsequential difference.

The experimental apparatus is sketched in Fig. \ref{setup2}. The PDC
source is a 5 mm long LiIO$_{3}$ bulk crystal, pumped by 400 nm
light, that produces pairs of correlated photons at 800 nm. The pump
light is obtained by doubling the frequency of the output of a
mode-locked laser (with a repetition rate of $\approx$80 MHz) pumped
by a 532 nm laser. Two interference filters (IF) with spectral
bandwidth full-width-half-maximum of 20 nm are placed in both the
heralded and heralding arms to reduce background light. Microscope
objectives (20x) collect the light into multi-mode fibers (MMF) and
the photons are finally counted by Si-Single-Photon Avalanche Diodes
(Si-SPADs) operating in Geiger mode. A half-wave plate ($\lambda/2$)
and polarizing beamsplitter (PBS) are used for our polarization
projective measurements.

To verify the single photon nature of our source, which is critical
for our test, we quantify the possibility of having more than one
photon in the heralded arm after detecting the heralding photon. For
this we use the same setup as for the main experiment (Fig. 1), but
we substitute into the heralded arm the multi-mode fiber with an
integrated 50:50 beam-splitter that sends the photons to two
Si-SPADs. The purity of a single-photon source can be described by
means of the two parameters $\gamma_{1}=
\theta(\mathrm{1})/\theta(\mathrm{0})$ and $\gamma_{2}=
\theta(\mathrm{2})/\theta(\mathrm{1})$, where $\theta(\mathrm{0})$,
$\theta(\mathrm{1})$, and $\theta(\mathrm{2})$ are the probabilities
of the heralded arm producing 0, 1, or 2 counts for each heralding
count, respectively.

In general, a heralding detection announces the arrival of a
``pulse" containing $n$ photons at the heralded channel. The
probability that neither of the Si-SPADs fire for a heralded optical
pulse containing $n$ photons is
\begin{eqnarray}\nonumber
\theta(\mathrm{0}|n)&=&\sum_{m=0}^{n} (1-\tau_A)^{m}~ (1-\tau_B)^{n-m}~  B(m|n;p=0.5) \\
&=& \left(1-\frac{\tau_A+\tau_B}{2} \right)^{n},
\end{eqnarray}
where $p$ represents the BS splitting ratio ($p=0.5$),
$B(m|n;p)=n![m!~ (n-m)!]^{-1}p^{m}(1-p)^{n-m}$ is the binomial
distribution representing the splitting of $n$ photons towards the
two Si-SPADs, and $\tau_A$ and $\tau_B$ are the detection
efficiencies of each Si-SPAD (that includes all collection and
optical losses in the detection channel). Analogously, the
probability of observing 1 or 2 counts due to a heralded optical
pulse with $n$ photons are respectively
\begin{eqnarray}\nonumber
 \theta(\mathrm{1}|n)&=& \sum_{m=0}^{n} \{  [1-(1-\tau_A)^{m}] (1-\tau_B)^{n-m} +  (1-\tau_A)^{m}[ 1-(1-\tau_B)^{n-m}]\} B(m|n;p=0.5)
 \\ \nonumber
 &=& \left(1-\frac{\tau_A}{2} \right)^{n} + \left(1-\frac{\tau_B}{2} \right)^{n} - 2 \left(1-\frac{\tau_A+\tau_B}{2} \right)^{n}, \text{ and} \\ \nonumber
\theta(\mathrm{2}|n)&=&\sum_{m=0}^{n}   [1-(1-\tau_A)^{m}] [ 1-(1-\tau_B)^{n-m}]  B(m|n;p=0.5) \\
 &=& 1- \left(1-\frac{\tau_A}{2} \right)^{n} - \left(1-\frac{\tau_B}{2} \right)^{n} + \left(1-\frac{\tau_A+\tau_B}{2} \right)^{n}.
\end{eqnarray}

From these we get $\theta(\mathrm{0})=\sum_{n}
\theta(\mathrm{0}|n)\mathcal{P}(n)$, $\theta(\mathrm{1})=\sum_{n}
\theta(\mathrm{1}|n)\mathcal{P}(n)$, and
$\theta(\mathrm{2})=\sum_{n} \theta(\mathrm{2}|n)\mathcal{P}(n)$ for
$\mathcal{P}(n)$ being the general probability distribution of the
number of photons in a heralded optical pulse. Assuming both
Si-SPADs have the same detection efficiencies ($\tau_A=\tau_B=
\tau$), we list in the Table I the parameters for an ideal
single-photon source ($\mathcal{P}(n)=\delta_{n,1}$) and a
Poissonian source ($\mathcal{P}(n)=\mu^{n} e^{-\mu}/n!$) with, on
average, $\mu$ photons per pulse. Comparing the measured results to
the the theoretical values for the two types of sources, we see that
while our source differs from an ideal single-photon source that
emits one photon per pulse, the very small $\gamma_{2} /
\gamma_{1}$, supports the point that conditional single-photon
output does dominate our source's output.

\begin{table}
\begin{center}
\begin{tabular}{ccccc} \hline \hline
 Source      & Poisson & Ideal single & without & background  \\
 Parameter   &    source & photon source & background subtracted &  subtracted \\
\hline
  &  &\\$\theta(\mathrm{0}) $ &    $\exp(-\tau \mu )$& $1-\tau $ &  &
  \\$\theta(\mathrm{1}) $  & $2 [\exp(-\tau \mu/2 )-\exp(-\tau \mu )]$&   $\tau$ &  &
  \\$\theta(\mathrm{2}) $  & $1- 2 \exp(-\tau \mu/2 ) + \exp(-\tau
\mu ) $&    0&   &
   \\$\gamma_1 $   & $2 [\exp(\tau \mu/2 )-1]$& $\tau/(1-\tau )$ & 0.0578(2)  & 0.0498(2) \\
$\gamma_2  $   & $ [\exp(\tau \mu/2 )-1]/2$& 0 & 0.0013(2)   & 0.0007(3)\\
$\gamma_{2} / \gamma_{1}  $   & 1/4& 0& 0.022(3)  & 0.015(5)
\\ 
  &  &    \\ \hline \hline
\end{tabular}
\end{center}

\caption{\label{tab:table1} Theoretical and measured photon source
parameters compared for an Ideal single-photon source and a Poisson
source. Standard uncertainties (shown in parenthesis) account for
both count fluctuations and random misalignment of the polarizers. }
\end{table}

The expectation values $\langle\widehat{A}\rangle$,
$\langle\widehat{B}\rangle$, $\langle\widehat{A}^{2}\rangle$,
$\langle\widehat{B}^{2}\rangle$  can be obtained experimentally by
projecting the heralded photons onto the linear polarizations states
as the operators $\widehat{A}$ and $\widehat{B}$ can be rewritten as
\begin{eqnarray}
\widehat{A}&=&a \widehat{P}_{0} \\
\widehat{B}&=&b\left(\frac{1+r}{2} \widehat{P}_{\frac{\beta}{2}}+
\frac{1-r}{2} \widehat{P}_{\frac{\beta+\pi}{2}} \right),
\end{eqnarray}
where $\widehat{P}_{\theta}$ is the projection operator on the state
$|s(\theta)\rangle= \cos \theta |H\rangle + \sin \theta |V\rangle$.

If we choose the parameter values $a=0.74$, $b=1.2987$,
 $r=3/5$, $\beta=2/9 ~ \pi $ (we note that, with this choice, the condition in Eq. (\ref{condvio}) is satisfied), and
$\psi=-11/36 ~ \pi$ the results for $\langle \widehat{B}^{2} \rangle
-\langle \widehat{A}^{2} \rangle $, and $\langle \widehat{B} \rangle
-\langle \widehat{A} \rangle$ predicted by quantum theory are those
presented in Table 2, while the minimum eigenvalue of $ \widehat{B}
- \widehat{A} $ is $d_{-} = 0.0189$ (satisfying the $ \widehat{B} -
\widehat{A} > 0$ requirement for the Alicki test), where
\begin{equation}\label{dminus}
d_{-}\equiv\frac{1}{2}\left\{b- a-\sqrt{a^2 +b^2~r^2- 2~ a~
b~r~\cos\beta } \right\}.
\end{equation}


\begin{table}
\begin{center}
\begin{tabular}{lccr} \hline \hline
Quantity & Measurement& Measurement &QM theory\\
 & (this work) & (previous work)\\
\hline
  &  &    \\$\langle \widehat{B}\rangle  - \langle
\widehat{A}\rangle $ & 0.0701(15)  & 0.058(11) & 0.0685\\
$\langle \widehat{B}^{2}\rangle - \langle
\widehat{A}^{2}\rangle  $ & -0.0461(10)  & -0.0403(66) & -0.0449\\
\\
deviation from  & 46.1 $\sigma$ & 9.4 $\sigma$\\
  classicality     \\ \hline \hline
\end{tabular}
\end{center}
\caption{\label{tab:table2} Non-classicality test experimental
results. Standard uncertainties (shown in parenthesis) account for
count fluctuations, random misalignment of the polarizers. The
contribution to the uncertainty budget due to non-ideal behavior of
our heralded single-photon source is negligible.}
\end{table}

The experimental results are presented in Table II. From the value
of $\langle \widehat{B}^{2}\rangle - \langle \widehat{A}^{2}\rangle
$ we see a very large violation, $\approx46$ standard deviations, of
the classical limit of $\langle \widehat{B}^{2}\rangle - \langle
\widehat{A}^{2}\rangle >0$.

\section{Conclusion}

In conclusion, we have presented a very simple and efficient
experimental implementation of Alicki's proposed nonclassicality
test, that results in a large (46.1 standard deviations) and low
noise violation of the Alicki classicality condition. This 5x
improvement over our previous result was achieved by moving from the
infrared to the visible, where we can take advantage of higher-efficiency 
and lower-noise photon detectors.

\section*{ Acknowledgments}

This work has been supported in part by Regione Piemonte (E14), San
Paolo Foundation, and by the MURI Center for Photonic Quantum
Information Systems (Army Research Office (ARO)/ Intelligence
Advanced Research Projects Activity (IARPA) program
DAAD19-03-1-0199), the IARPA entangled source programs. N. G. was
partially supported by the Swiss NCCR-QP.

\end{document}